# Observation of quantum Hawking radiation and its entanglement in an analogue black hole

Jeff Steinhauer

Department of Physics, Technion—Israel Institute of Technology, Technion City, Haifa 32000, Israel

**We observe spontaneous Hawking radiation, stimulated by quantum vacuum fluctuations, emanating from an analogue black hole in an atomic Bose-Einstein condensate. Correlations are observed between the Hawking particles outside the black hole and the partner particles inside. These correlations indicate an approximately thermal distribution of Hawking radiation. We find that the high energy pairs are entangled, while the low energy pairs are not, within the reasonable assumption that excitations with different frequencies are not correlated. The entanglement verifies the quantum nature of the Hawking radiation. The results are consistent with a driven oscillation experiment and a numerical simulation.**

50 years ago, Bekenstein discovered the field of black hole thermodynamics[1]. This field has vast and deep implications, far beyond the physics of black holes themselves. The most important prediction of the field is that of Hawking radiation[2,3]. By making an approximation to the still-unknown laws of quantum gravity, Hawking predicted that the horizon of the black hole should emit a thermal distribution of particles. Furthermore, each Hawking particle should be entangled with a partner particle falling into the black hole. This presents a puzzle of information loss, and even the unitarity of quantum mechanics falls into question[4-6].

Despite the importance of black hole thermodynamics, there were no experimental results to provide guidance. The problem is that the Hawking radiation emanating from a real black hole should be exceedingly weak. To facilitate observation, Unruh suggested that an analogue black hole can be created in the laboratory, where sound plays the role of light, and the local flow velocity and speed of sound determine the metric of the analogue spacetime[7]. Nevertheless, thermal Hawking radiation had never been observed before this work.

Since the idea of analogue Hawking radiation was presented[7], there has been a vast theoretical investigation of a variety of possible analogue black holes[8-21]. It was predicted that the Hawking radiation could be observed by the density correlations between the Hawking and partner particles[9,10]. The entanglement of the Hawking pairs has also been studied theoretically[22-28]. Recently, we explained that the density correlations could be used to observe the entanglement[24], and we have implemented our technique here.

Over the past several years, we have systematically prepared for the observation of thermal Hawking radiation by studying analogue black hole creation[29], phonon propagation[30], thermal distributions of phonons[31], and self-amplifying Hawking radiation[32]. Our observation of Hawking radiation is performed in a Bose-Einstein condensate, a system in which the quantum vacuum fluctuations are guaranteed by the underlying pointlike atoms composing the condensate. There are experiments in several other systems currently underway with the hopes of observing Hawking radiation[33-37]. Furthermore, stimulated classical mode mixing at a white hole horizon was observed[38,39].

It has been suggested that the Hawking and partner particles can be observed by studying the 2-body correlation function between points on opposite sides of the horizon[9,10,12,40]. The correlation function is given by $G^{(2)}(x,x') = \sqrt{n_{out}n_{in}\xi_{out}\xi_{in}}\langle\delta n(x)\delta n(x')\rangle/n_{out}n_{in}$, where $n(x)$ is the 1D density of the condensate forming the black hole, and $n_{out}$ and $n_{in}$ are the average densities outside and inside the black hole, respectively. The position $x$ is in units of the shortest length scale of the condensate $\xi \equiv \sqrt{\xi_{out}\xi_{in}}$, where $\xi_{out}$ and $\xi_{in}$ are the healing lengths outside and inside the black hole respectively, and $\xi_i = \hbar/mc_i$, where $c_i$ is the speed of sound and $m$ is the mass of an atom in the condensate. The strength of the fluctuations are characterized by the prefactor $\sqrt{n_{out}n_{in}\xi_{out}\xi_{in}}$; the lower the number, the larger the signal of Hawking radiation[10]. Fig. 1a shows the theoretical $G^{(2)}$ in vacuum, in the hydrodynamic limit of low Hawking temperature in which dispersion can be neglected, in strict analogy with real gravity[9]. Correlations are seen along the line of equal propagation times from the horizon, outside and inside the black hole. These are the correlations between the Hawking and partner particles. Such correlations should also exist in a real black hole, within Hawking's approximation[40].

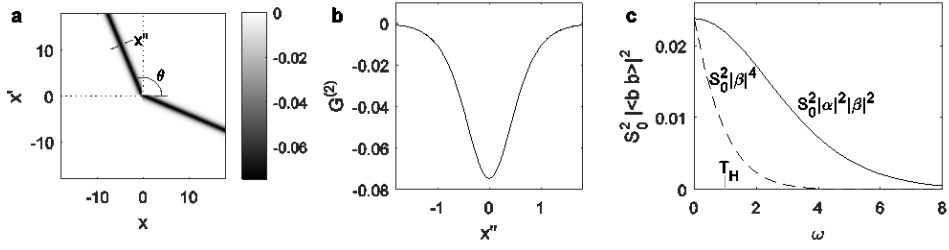

**Figure 1 | Hawking radiation and entanglement in the gravitational analogy. a**, Density-density correlations between the Hawking and partner particles[9]. Positive (negative) $x$ or $x'$ corresponds to inside (outside) the black hole. **b**, The profile of **a** in the $x''$ direction. **c**, The entanglement. The solid curve is proportional to the squared magnitude of the Fourier transform of **b**. The dashed curve is proportional to the population squared.

We find that much insight can be gained by considering the Fourier transform of individual quadrants of $G^{(2)}$ (Ref. 24). Most importantly, the Fourier transform of the correlations between points outside and inside the black hole (the quadrant outlined with dotted lines in Fig. 1a) gives the $k$-space correlation spectrum $\langle \hat{b}_{k_{HR}} \hat{b}_{k_P} \rangle$, where $\hat{b}_{k_{HR}}$ is the annihilation operator for a Hawking particle with wavenumber $k_{HR}$ localized outside the black hole, and $\hat{b}_{k_P}$ is the annihilation operator for a partner particle localized inside the black hole[24]. Specifically,

$$S_0 \langle \hat{b}_{k_{HR}} \hat{b}_{k_P} \rangle = \sqrt{\frac{\xi_{\text{out}} \xi_{\text{in}}}{L_{\text{out}} L_{\text{in}}}} \int dx dx' e^{ik_{HR}x} e^{ik_P x'} G^{(2)}(x,x') \qquad (1)$$

where $S_0 \equiv \left(U_{k_{HR}} + V_{k_{HR}}\right)\left(U_{k_P} + V_{k_P}\right)$ and $U_i$ and $V_i$ are the Bogoliubov coefficients for the phonons, which are completely determined by $\xi_i k_i$. The length of each region is given by $L_i$. The coordinates $x$ and $x'$ are integrated over the intervals $[-L_{\text{out}}, 0]$ and $[0, L_{\text{in}}]$, respectively. If the correlation feature is elongated with a constant cross-section as in Fig. 1a, then (1) reduces to

$$S_0 \langle \hat{b}_{k_{HR}} \hat{b}_{k_P} \rangle = \sqrt{-\tan\theta - \cot\theta} \int dx'' e^{ikx''} G^{(2)}(x,x') \tag{2}$$

where $x''$ is the coordinate perpendicular to the feature in units of $\xi$, $\theta$ is the angle of the correlation feature in the $x$-$x'$ plane as in Fig. 1a, $k$ is in units of $\xi^{-1}$, $k_{HR} = k\sin\theta$, and $k_P = -k\cos\theta$. In (1), we have neglected the quantities $\langle \hat{b}^\dagger_{-k_{HR}} \hat{b}^\dagger_{-k_P} \rangle$, $\langle \hat{b}^\dagger_{-k_{HR}} \hat{b}_{k_P} \rangle$, and $\langle \hat{b}^\dagger_{-k_P} \hat{b}_{k_{HR}} \rangle$. These terms all involve negative $k_i$, corresponding to phonons directed with the flow in the comoving frame. The frequencies of these phonons are increased by $2|k_i|v_i$ relative to the positive-$k$ phonons, due to the Doppler shift. Thus, the neglected terms represent correlations between widely separated phonons on opposite sides of the horizon with different frequencies. Hawking radiation would not create such correlations, and it is not clear what would. Thus, (1) should be a good approximation in the general case of spatially separated regions in an inhomogeneous, flowing condensate. The reasonable assumption that the terms can be neglected allows for our measurement of the entanglement.

As we explained in our recent work[24], the density-density correlation function can be used to measure the entanglement of the Hawking and partner particles. The narrow width of the correlation band implies that $\langle \hat{b}_{k_{HR}} \hat{b}_{k_P} \rangle$ is broad in $k$-space. If it is even broader than the thermal population distribution of the Hawking radiation, then the state is nonseparable, i.e. entangled. Specifically, we would like to evaluate the nonseparability measure $\Delta$ (Ref. 22), which is equivalent to the Peres-Horodecki criterion[22,24].

$$\Delta \equiv \langle \hat{b}^\dagger_{k_{HR}} \hat{b}_{k_{HR}} \rangle \langle \hat{b}^\dagger_{k_P} \hat{b}_{k_P} \rangle - \left| \langle \hat{b}_{k_{HR}} \hat{b}_{k_P} \rangle \right|^2 \tag{3}$$

If $\Delta$ is negative, then the state is nonseparable. The first term in (3) is the product of the populations of Hawking and partner particles, which should be the same. All of our measurements are density measurements which commute with one another. Nevertheless, we evaluate the non-commuting terms in (3) by making the assumption that we can neglect the correlations between phonons of different frequencies.

We can verify the technique by applying it to the hydrodynamic limit. By (2), the Fourier transform of Fig. 1b gives $S_0^2|\langle \hat{b}_{k_{HR}} \hat{b}_{k_P}\rangle|^2$, as indicated by the solid curve in Fig. 1c. This curve precisely agrees with the expression for a linear dispersion relation in vacuum[11] $|\langle \hat{b}_{k_{HR}} \hat{b}_{k_P}\rangle|^2 = |\alpha|^2|\beta|^2$, where $\alpha$ and $\beta$ are Bogoliubov coefficients, $|\alpha|^2 = |\beta|^2 + 1$, and the population is $|\beta|^2 = (e^{\hbar\omega_k/k_B T_H} - 1)^{-1}$. The first term of (3) is $|\beta|^4$, which is always smaller than the second term as seen in Fig. 1c. In fact, the second term is equal to the largest value allowed by the Heisenberg uncertainty principle, $|\beta|^2(|\beta|^2 + 1)$ (Ref. 22). In this sense, the Hawking pairs in the hydrodynamic limit are maximally entangled at all frequencies. However, the ratio between the two terms in (3) approaches unity at long wavelengths, as seen in Fig. 1c. Thus, the entanglement is most easily destroyed there. This is consistent with the results of Refs. 22, 23, 27, and 28, which found a loss of entanglement at low $k$ due to dispersion and/or finite temperatures.

**The experimental system**

Fig. 2a shows the Bose-Einstein condensate of $^{87}$Rb atoms confined radially by a narrow laser beam (3.6 μm waist, 812 nm wavelength). The radial trap frequency of 140 Hz is greater than the maximum interaction energy of 70 Hz so the behavior is 1D, which allows the phonons to propagate with little decay, as we studied in Ref. 30. The horizon is hydrodynamic in the sense that its width is a few $\xi$, as shown in the inset of Fig. 2b. It is seen that the density profile approximates half a gray soliton[12]. On the other hand, this relatively steep density gradient at the horizon maximizes the Hawking temperature, and therefore results in an observable amount of Hawking radiation. The horizon is created by a very sharp potential step, achieved by short-wavelength laser light (0.442 µm), and high-resolution optics (NA 0.5). The step in the external potential is thus narrow compared to $\xi = 2.0$ µm. The step potential is swept along the condensate at a constant speed of $0.18$ mm sec$^{-1}$, as indicated by the black line in Fig. 2d. Deep in the outside region to the left of the step, the condensate is at rest. The flow velocity increases as the step is approached. To the right of the step, the condensate flows at supersonic speed due to the potential drop. It is useful to consider the horizon frame in which the step is at rest at the origin. In this frame, in the outside region, the condensate flows from left to right at the speed $v_{out} = 0.24$ mm sec$^{-1}$, which is close to the applied speed. This flow is subsonic since it is less than the speed of sound $c_{out} = 0.57$ mm sec$^{-1}$, so phonons can travel against the flow and escape the black hole. In contrast, inside the black hole, the flow is supersonic, at speed $v_{in} = 1.02$ mm sec$^{-1}$, which is greater than the speed of sound $c_{in} = 0.25$ mm sec$^{-1}$. Here, the phonons are trapped in that they cannot reach the horizon, in analogy with photons inside a black hole. It is seen that $c_{in}$ is close to $v_{out}$, as expected for a gray soliton[12]. In other words, the green soliton curve in the inset of Fig. 2b agrees with the experimental curve for x > 0 with no free parameters. These values of $v_{out}$, $c_{out}$, $v_{in}$, and $c_{in}$ are determined below by fitting the Doppler-shifted Bogoliubov dispersion relation to the measured dispersion relation.

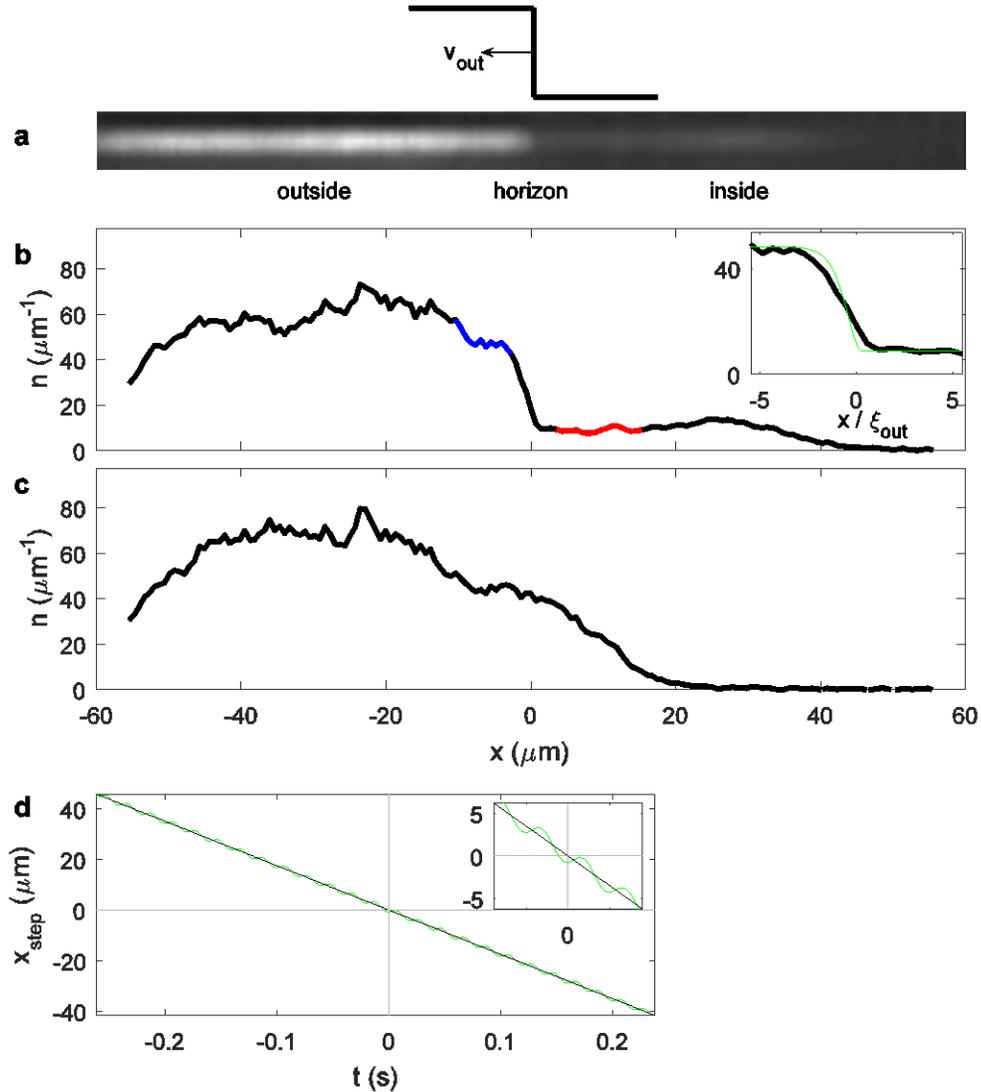

**Figure 2 | The analogue black hole. a,** The 1D Bose-Einstein condensate, which traps phonons in the region to the right of the horizon. The average over the ensemble is shown. **b,** The profile of **a**. Blue and red indicate the regions outside and inside the black hole where the Hawking/partner correlations are observed. The green curve in the inset is half a gray soliton[12]. **c,** The profile of the condensate without the step potential, obtained from an average of 1400 images. **d,** The position of the step potential as a function of time. The black line is used for observing Hawking radiation. The green curve is used for generating waves.

A wavelength of 0.78 μm is employed for the phase contrast imaging of the black hole. This infrared wavelength is far from the blue wavelength of the step potential, so the imaging is performed through an additional NA 0.5 lens which views the condensate along an axis perpendicular to that of the lens of the potential.

**Oscillating horizon experiment**

In order to study the properties of the analogue black hole, a preliminary experiment is performed in which the horizon is caused to oscillate at a definite frequency. This generates correlated waves traveling inward and outward from the horizon. The experiment is repeated 200 to 1200 times for each frequency, with a different phase each repetition. The resulting correlation function for a given frequency shows a wave pattern, giving the wavelength as a function of position (Fig. 3a). It also shows the location and sharpness of the horizon, indicated by a dashed line.

The green rectangle in Fig. 3a corresponds to the spatial region where Hawking/partner correlations are seen in Fig. 4a. By computing the 2D Fourier transform within the rectangle as shown in Fig. 3b, $k_{\text{out}}$ and $k_{\text{in}}$ are determined for each applied frequency, as well as the full-width-half-max (FWHM) of the $k$-distribution. The resulting dispersion relation is shown in Figs. 3d,e. If two branches are visible and resolved inside the black hole for a given frequency, then they are both shown. For most points, the FWHM reflects the dimensions of the rectangle in Fig. 3a. The exception occurs at $k_{\text{peak}}$, where the wavepacket in $k$-space is significantly broader than the FWHM determined by the rectangle. This point has zero group velocity and diverging density of states per unit frequency, so the finite frequency width due to the finite duration of the experiment excites a broad wavepacket in $k$-space. Although the center of the outgoing wavepacket must be below $k_{\text{peak}}$, components above $k_{\text{peak}}$ are expected. Indeed, the group velocity concept applies to wavepackets rather than single modes. Such an outgoing wavepacket is clearly possible when viewed in the comoving frame in which $k_{\text{peak}}$ is not a special point. The $k$-content of the outgoing wavepacket excited by $\omega_{\text{peak}}$ is seen in Fig. 3b. The components above $k_{\text{peak}}$ (above the dotted line) are clearly visible. We can define $k_{\text{max}}$ as the highest observed $k$ in the outgoing wavepacket, as seen in Fig. 3c. The components above $k_{\text{peak}}$ are also clear in Fig. 3f which shows a numerical simulation using the one-dimensional Gross-Pitaevskii equation, with parameters similar to the experiment.

The measured strength of the stimulated correlations is consistent with the assumption that the oscillating horizon is equivalent to incoming particles. In this case, the number of particles produced with the same energy as the incoming particles is proportional to $|\beta|^2 + 1$, and the number produced with opposite energy is proportional to $|\beta|^2$. The value of $|\beta|^2$ is taken to be the population of spontaneous Hawking radiation measured in the main experiment. Furthermore, for the general case of excitation by a weak probe, the rate of particle production is proportional to $S_0$ (Ref 41). Some examples are neutron scattering from a superfluid[42], and scattering of a Bose-Einstein condensate by a spatially oscillating potential with a narrow[43] or broad[30] frequency spectrum. Applying this to the oscillating horizon, and assuming that every frequency has the same correlation properties so $|\langle \hat{b}_{k_{\text{out}}} \hat{b}_{k_{\text{in}}} \rangle|^2 \propto \langle \hat{b}^\dagger_{k_{\text{out}}} \hat{b}_{k_{\text{out}}} \rangle \langle \hat{b}^\dagger_{k_{\text{in}}} \hat{b}_{k_{\text{in}}} \rangle$, one obtains $S_0^2 |\langle \hat{b}_{k_{\text{out}}} \hat{b}_{k_{\text{in}}} \rangle|^2 \propto S_0^4 |\beta|^2 (|\beta|^2 + 1)$. This expression agrees well with the measurement, as indicated by the dashed curve in Fig. 3h, which suggests that both the

spontaneous experiment and the driven experiment are governed by the same Hawking temperature.

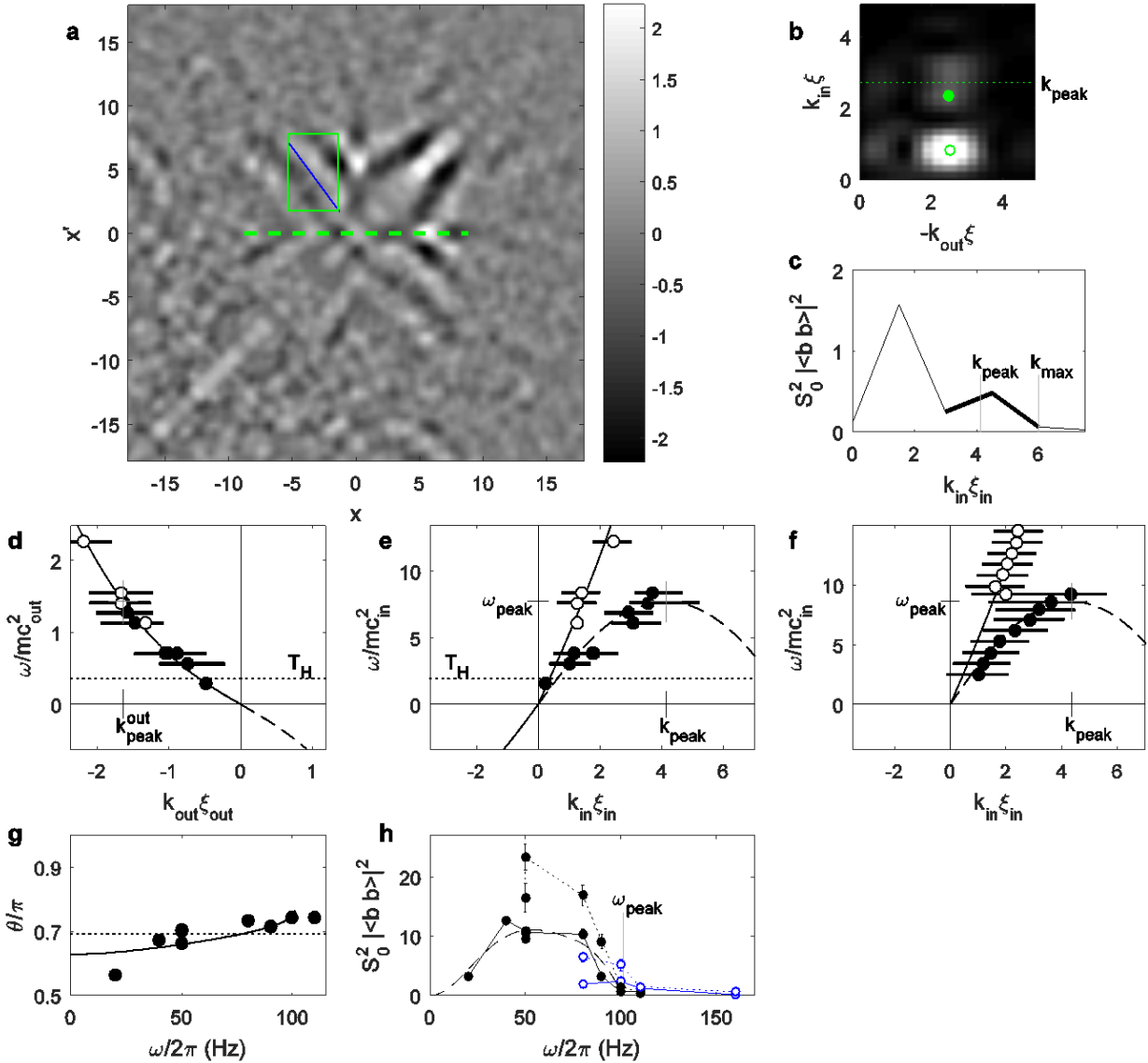

**Figure 3 | Oscillating horizon experiment.** A preliminary experiment is shown in which the step potential at the horizon is caused to oscillate at a definite frequency with an amplitude of 1 µm. **a,** The correlation function for 50 Hz. The blue line indicates the line of equal propagation times on opposite sides of the horizon, determined by the Fourier transform. **b,** The 2D Fourier transform computed within the green rectangle of **a**. $\omega_{peak}$ = 100 Hz is shown. The dotted line

indicates $k_\text{peak}$. The filled and open circles are as in **d,e**. The correlation function is padded with zeros before the Fourier transform. **c,** The profile of b. No padding is used. **d,e,** The dispersion relation outside and inside the black hole. The filled (open) circles are directed toward (away from) the horizon inside the black hole in the comoving frame. The horizontal bars indicate the FWHM of the outgoing wavepacket. The curves are fits of the Bogoliubov dispersion relation including the Doppler shift. The dashed curves are the negative norm branches. The Hawking temperature is indicated by the dotted line. **f,** The simulated dispersion relation inside the black hole. The dashed and solid curves are determined from the parameters of the simulation rather than from a fit. **g,** The angle of equal propagation times $\tan^{-1}(k_\text{in}/k_\text{out})$ (such as the angle of the blue line in **a**), as a function of the applied frequency. The solid curve is the angle from the Bogoliubov dispersion relation. The dotted line indicates the measured angle of the Hawking/partner correlations in Fig. 4a. **h,** The correlations of the preliminary experiment. The solid curves indicate $S_0^2 |\langle \hat{b}_{k_\text{out}} \hat{b}_{k_\text{in}} \rangle|^2$. The dotted curves show $S_0^2 \langle \hat{b}^\dagger_{k_\text{out}} \hat{b}_{k_\text{out}} \rangle \langle \hat{b}^\dagger_{k_\text{in}} \hat{b}_{k_\text{in}} \rangle$. The filled and open circles are as in **d,e**. The dashed curve is proportional to $S_0^4 |\beta|^2 (|\beta|^2 + 1)$. The error bars indicate the standard error of the mean.

**Observation of Hawking radiation**

Fig. 4a shows the measured correlation function between pairs of points $(x, x')$ along the analogue black hole. This correlation function is computed from an ensemble of 4600 repetitions of the experiment, requiring 6 days of continuous measurement. Fig. 4a has been filtered to remove the effects of imaging shot noise at high frequencies, imaging fringes, and overall slopes due to the profile of the analogue black hole. This includes smoothing the diagonal region shaded in blue.

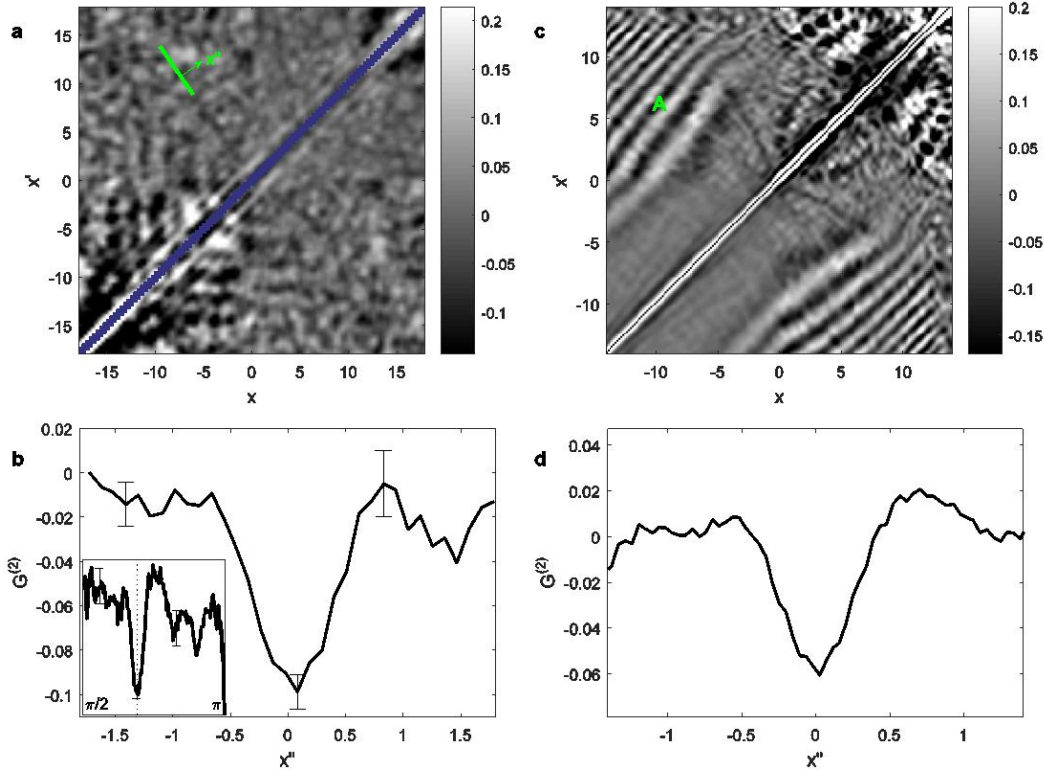

**Figure 4 | Observation of Hawking/partner pairs. a,** The two-body correlation function. The horizon is at the origin. The dark bands emanating from the horizon are the correlations between the Hawking and partner particles. The green line indicates the angle found in the inset to **b**. **b,** The profile of the Hawking-partner correlations. The Fourier transform of this curve measures the entanglement of the Hawking pairs. The error bars indicate the standard error of the mean. The inset shows the angular profile of **a**. **c,** Numerical simulation. The fringes marked "A" are an artifact of the creation of the fluctuations. **d,** The profile of the simulation.

The upper left and lower right quadrants of Fig. 4a show the correlations between points on opposite sides of the horizon. In contrast to Fig. 3a, the broad Hawking spectrum gives correlations only for points with equal propagation times. The dark band of these points is clearly visible emanating from the horizon. This is the observation of the Hawking and partner particles. The correlation band is relatively straight which implies that there is not much spatial variation in the propagation speed in the blue and red regions of Fig. 2b. Indeed, the density varies by only 8% and 10% RMS in the blue and red regions, respectively.

To determine the angle of the correlation band, the correlation function is averaged along radial line segments at various angles, and with lengths equal to that of the correlation band. The result is shown in the inset to Fig. 4b. The minimum is clearly seen at the angle indicated by the dotted line, and a solid green line in Fig. 4a. This angle is also indicated by a dotted line in Fig. 3g, where good agreement is seen with the range of angles produced in the preliminary experiment. The dotted line is seen to fall near the middle of the range, rather than the angle corresponding to $\omega = 0$. A similar shift from the $\omega = 0$ angle is apparent in the simulations of Ref. 10.

The profile of the Hawking/partner correlation band is computed by averaging over its length, along lines parallel to the green line in Fig. 4a. This gives the profile in the $x''$ direction, as shown in Fig. 4b. For computing the profile, the correlation function is only filtered for high spatial frequencies beyond the resolution of the imaging system, and for imaging fringes. The finite area of the profile is the Fourier transform in (2), evaluated at small $k$. Since $S_0 \propto \omega$ in this limit, one finds $|\langle \hat{b}_{k_{HR}} \hat{b}_{k_P} \rangle|^2 \propto 1/\omega^2$. In the approximation of linear dispersion, $|\langle \hat{b}_{k_{HR}} \hat{b}_{k_P} \rangle|^2 \propto |\beta|^4$. Thus, the finite cross-section implies that the Hawking distribution at low energies is thermal in the sense that the population goes like $1/\omega$.

For comparison, Figs. 4c,d shows the a numerical simulation. In the simulation, we employ our short-pulse Bragg scattering technique[30] to introduce a spectrum of fluctuations. Specifically, a random potential is chosen for each run. The potential is filtered with the profile $(U_k + V_k)^{-1}$ so that it will produce an approximately correct zero-temperature distribution. The potential is turned on for a very short time to create the fluctuations. The fluctuations are introduced well after the creation of the horizon, demonstrating that the Hawking radiation occurs in an approximately stationary background. The Hawking/partner correlation feature in the simulation is seen to be in good agreement with the experimental Figs. 4a,b.

We can also compare the Hawking/partner correlations with the simulation of Ref. 10. There, the width of the correlation band was several times wider than our measurement. This can be explained at least partially by $k_{\text{peak}}$ which was several times smaller.

**Measuring the population of the Hawking radiation**

In preparation for observing Hawking radiation in our analogue system, we developed a technique of observing the real and virtual phonons[31,32]. Fig. 5a shows the power spectrum (the static structure factor $S(k)$) of the density fluctuations in the spatial region outside the black hole where the Hawking radiation is observed in Fig. 4a. The measured $S(k)$ is seen to decrease for decreasing $k$, which indicates that quantum fluctuations dominate. The theoretical spectra at various temperatures are also shown, in which the Planck distribution is brought linearly to zero[11-13] at the measured $\omega_{\text{peak}}$ of Fig. 3e, as shown in the inset of Fig. 5b. The best fit gives a measured Hawking temperature of $k_B T_H = 0.36\ mc_{\text{out}}^2$, slightly above the range of predicted approximate maximum values $0.25 - 0.32\ mc_{\text{out}}^2$ (Refs. 11, 12, and 29). The fit includes an overall multiplicative factor of 2.2 which is applied to the data, since $S(k)$ must approach unity at high energies due to the pointlike particles composing the condensate. The reduced sensitivity implied by the factor is likely due to aberrations in the coherent-light imaging system. This factor is independent of $k$, as measured in Ref. 30 for similar optics. A very similar factor of 2.3 is found inside the black hole. These factors are also applied to the fluctuations in Figs. 3, 4, and 6. The population is given by the relative difference between the measured and zero-temperature curves of Fig. 5a, as shown in Fig. 5b. This includes the Hawking radiation as well as any background phonons. The measured population is seen to agree well with the theoretical distribution of Hawking radiation for $k_\text{B} T_\text{H} = 0.36 \, mc_\text{out}^2$.

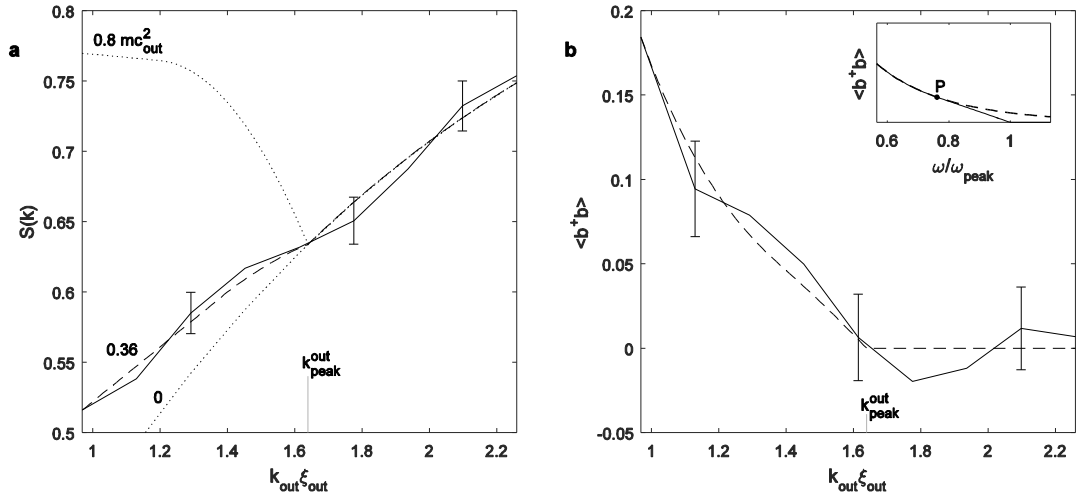

**Figure 5 | The measured population of the Hawking radiation. a,** The power spectrum of the real and virtual fluctuations. The solid curve is the measurement. The dashed curve is the best-fit theoretical spectrum at $k_\text{B} T_\text{H} = 0.36 \, mc_\text{out}^2$. The dotted curves show other temperatures for comparison. The vertical line is the measured $k_\text{peak}^\text{out}$ from Fig. 3d. The error bars indicate the standard error of the mean. **b,** The phonon population derived from **a**. The dashed curve

indicates $k_B T_H = 0.36\ mc_{out}^2$. The inset shows the Planck distribution (dashed curve) brought linearly to zero at $\omega_{peak}$. The point $P$ is chosen to give a continuous first derivative.

**Measuring the entanglement of the Hawking pairs**

By (2), we can extract $S_0^2 |\langle \hat{b}_{k_{HR}} \hat{b}_{k_P} \rangle|^2$ as a function of $k_P$ from Fig. 4b, as indicated by the solid curve in Fig. 6a. For the larger $k$-values, this observed spectrum is in agreement with the light gray $S_0^2 |\beta|^2 (|\beta|^2 + 1)$ curve, which indicates that the Hawking radiation is thermal at these high energies. The theoretical curve has been convoluted with the $k$-distribution of the outgoing modes near $k_{peak}$, which is indicated in Fig. 3c by a thick line. The components above $k_{peak}$ in the observed spectrum cause a narrowing of the correlation feature in Figs. 4a,b. For lower $k$ however, the observed spectrum is far below the theoretical curve. This implies that the long-wavelength Hawking pairs are less correlated than expected, or they are produced in smaller quantities than expected by a factor of approximately 3.6. Perhaps the low frequency waves did not have sufficient time to form. The suppression of the low $k$'s results in a narrower and shallower correlation feature in Figs. 4a,b. It may also explain the shift in the correlation feature from the $\omega = 0$ angle. The width of the observed spectrum relative to $k_{peak}$ and $k_{max}$ is qualitatively similar to the simulation shown in Fig. 6c.

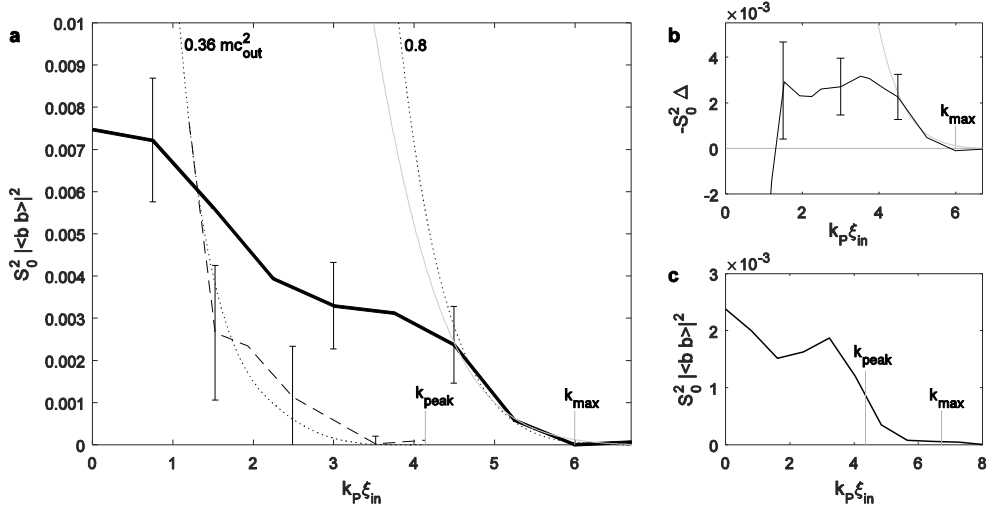

**Figure 6 | Observation of entanglement between the Hawking pairs. a,** The correlations. The solid curve indicates the correlations between the Hawking and partner particles. It is proportional to the square of the Fourier transform of Fig. 4b. The dashed curve is the measured population squared from Fig. 5b. The solid curve exceeding the dashed curve corresponds to entanglement. The dotted curves show the theoretical population squared at the observed value of $k_\mathrm{B} T_\mathrm{H} = 0.36\ mc_\mathrm{out}^2$, as well as at $0.8\ mc_\mathrm{out}^2$. The latter has been convoluted with the $k$-distribution of outgoing modes near $k_\mathrm{peak}$. The solid gray curve indicates the maximally entangled Heisenberg limit for $0.36\ mc_\mathrm{out}^2$. The measured $k_\mathrm{max}$ and $k_\mathrm{peak}$ are from Figs. 3c,e. The error bars indicate the standard error of the mean. **b,** The nonseparability measure, the difference between the solid and dashed curves of **a**. Positive values of $-\Delta$ correspond to entanglement. The error bars shown are sufficiently spaced to be statistically independent. **c,** The numerical simulation corresponding to **a**. The curve is proportional to the square of the Fourier transform of Fig. 4d. $k_\mathrm{max}$ and $k_\mathrm{peak}$ are from the simulation of Fig. 3f.

The measured population squared from Fig. 5b is indicated in Fig. 6a by the dashed curve. The $S_0^2 |\langle \hat{b}_{k_{HR}} \hat{b}_{k_P} \rangle|^2$ solid curve exceeding the dashed curve indicates entanglement, by (3). It is seen that the high frequencies are entangled, whereas the low frequencies are not. Subtracting the curves gives $-\Delta$ as shown in the inset, where positive values indicate entanglement. A substantial $k$-range of entanglement is seen. For all of this range, $S_0^2 |\langle \hat{b}_{k_{HR}} \hat{b}_{k_P} \rangle|^2$ lies within the maximally entangled Heisenberg limit indicated by the solid gray curve, as expected. The theoretical prediction for this type of system would be near-maximal entanglement for all but the longest wavelengths[28]. Approximately maximal entanglement is indeed observed for the larger $k$-values. However, the entanglement is well below maximal for lower $k$ due to the suppression of the long wavelengths.

The probability of no entanglement is small. Firstly, the $0.8\ mc_{\text{out}}^2$ dotted curve in Fig. 6a indicates a temperature which would have resulted in a substantially reduced entanglement region, if it had been observed. However, such a high temperature is seen to be ruled out by the measurement in Fig. 5a. Secondly, if $S_0^2 |\langle \hat{b}_{k_{HR}} \hat{b}_{k_P} \rangle|^2$ in Fig. 6a were narrower by a factor of 0.53, then there would have been no entanglement. This would require that the profile of Fig. 4b would be wider by 90%, but the uncertainty in the width is only 16% as determined by a least-squares fit of a Gaussian including the contribution of the experimental error bars. In contrast to the entanglement seen for the Hawking radiation, the oscillating horizon experiment shows classical correlations ($\Delta > 0$) as expected (Fig. 3h).

In conclusion, Hawking radiation stimulated by quantum vacuum fluctuations has been observed in a quantum simulator of a black hole. Thermal behavior is seen at very low and very high energies. The measurements show that the experiment is well within the quantum regime, since the Hawking temperature determined from the population distribution is 1.2 nK, far below the measured 2.7 nK upper limit for quantum entanglement. A broad energy spectrum of entangled Hawking pairs is observed. The entanglement decreases for decreasing energy. The measurement reported here verifies Hawking's calculation, which is viewed as a milestone in the quest for quantum gravity. The observation of Hawking radiation and its entanglement confirms important elements in the discussion of information loss in a real black hole.

**Acknowledgments**

I thank Renaud Parentani, William Unruh, Florent Michel, Nicolas Pavloff, and Alessandro Fabbri for helpful comments. This work was supported by the Israel Science Foundation.

The data that support the plots within this paper and other findings of this study are available from the author upon reasonable request.

1. Bekenstein, J. D. Black holes and entropy. *Phys. Rev. D* **7**, 2333-2346 (1973).
2. Hawking, S. W. Black hole explosions? *Nature* **248**, 30-31 (1974).
3. Hawking, S. W. Particle creation by black holes. *Commun. Math. Phys.* **43**, 199-220 (1975).
4. Hawking, S. W. Breakdown of predictability in gravitational collapse. *Phys. Rev. D* **14**, 2460 (1976).
5. Susskind, L. The paradox of quantum black holes. *Nature Phys.* **2**, 665 (2006).
6. Almheiri, A., Marolf, D., Polchinski, J. & Sully, J. Black holes: complementarity or firewalls? *J. High Energy Phys.* **62** (2013).
7. Unruh, W. G. Experimental black-hole evaporation? *Phys. Rev. Lett.* **46**, 1351-1353 (1981).

8. Garay, L. J., Anglin, J. R., Cirac, J. I. & Zoller, P., Sonic analog of gravitational black holes in Bose-Einstein condensates. *Phys. Rev. Lett.* **85**, 4643-4647 (2000).

9. Balbinot, R., Fabbri, A., Fagnocchi, S., Recati, A. & Carusotto, I. Nonlocal density correlations as a signature of Hawking radiation from acoustic black holes. *Phys. Rev. A* **78**, 021603(R) (2008).

10. Carusotto, I., Fagnocchi, S., Recati, A., Balbinot, R. & Fabbri, A. Numerical observation of Hawking radiation from acoustic black holes in atomic Bose–Einstein condensates. *New J. Phys.* **10**, 103001 (2008).

11. Macher, J. & Parentani, R. Black-hole radiation in Bose-Einstein condensates. *Phys. Rev. A* **80**, 043601 (2009).

12. Larré, P.-É., Recati, A., Carusotto, I. & Pavloff, N. Quantum fluctuations around black hole horizons in Bose-Einstein condensates. *Phys. Rev. A* **85**, 013621 (2012).

13. Recati, A., Pavloff, N. & Carusotto, I. Bogoliubov theory of acoustic Hawking radiation in Bose-Einstein condensates. *Phys. Rev. A* **80**, 043603 (2009).

14. Barceló, C., Liberati, S. & Visser, M. Analogue gravity from Bose-Einstein condensates. *Class. Quant. Grav.* **18**, 1137-1156 (2001).

15. Corley S. & Jacobson, T. Black hole lasers. *Phys. Rev. D* **59**, 124011 (1999).

16. Jacobson T. A. & Volovik, G. E. Event horizons and ergoregions in $^3$He. *Phys. Rev. D* **58**, 064021 (1998).

17. Schützhold, R. & Unruh, W. G. Hawking radiation in an electromagnetic waveguide? *Phys. Rev. Lett.* **95**, 031301 (2005).

18. Giovanazzi, S. Hawking radiation in sonic black holes. *Phys. Rev. Lett.* **94**, 061302 (2005).

19. Horstmann, B., Reznik, B., Fagnocchi, S. & Cirac, J. I. Hawking radiation from an acoustic black hole on an ion ring. *Phys. Rev. Lett.* **104**, 250403 (2010).

20. Elazar, M. Fleurov, V. & Bar-Ad, S. All-optical event horizon in an optical analog of a Laval nozzle. *Phys. Rev. A* **86**, 063821 (2012).

21. Solnyshkov, D. D., Flayac, H. & Malpuech, G. Black holes and wormholes in spinor polariton condensates. *Phys. Rev. B* **84**, 233405 (2011).

22. Busch, X. & Parentani, R. Quantum entanglement in analogue Hawking radiation: When is the final state nonseparable? *Phys. Rev. D* **89**, 105024 (2014).

23. Finazzi, S. & Carusotto, I. Entangled phonons in atomic Bose-Einstein condensates. *Phys. Rev. A* **90**, 033607 (2014).

24. Steinhauer, J. Measuring the entanglement of analogue Hawking radiation by the density-density correlation function. *Phys. Rev. D* **92**, 024043 (2015).

25. de Nova, J. R. M., Sols, F. & Zapata, I. Violation of Cauchy-Schwarz inequalities by spontaneous Hawking radiation in resonant boson structures. *Phys. Rev. A* **89**, 043808 (2014).

26. Doukas, J. Adesso, G. & Fuentes, I. Ruling out stray thermal radiation in analogue black holes. arXiv 1404.4324.

27. Boiron, D., Fabbri, A., Larré, P.-É., Pavloff, N., Westbrook, C. I. & Ziń, P. Quantum signature of analog Hawking radiation in momentum space. *Phys. Rev. Lett.* **115**, 025301 (2015).

28. de Nova, J. R. M., Sols, F. & Zapata, I. Entanglement and violation of classical inequalities in the Hawking radiation of flowing atom condensates. *New J. Phys.* **17**, 105003 (2015).

29. Lahav, O., Itah, A., Blumkin, A., Gordon, C., Rinott, S., Zayats, A. & Steinhauer, J. Realization of a sonic black hole analog in a Bose-Einstein condensate. *Phys. Rev. Lett.* **105**, 240401 (2010).

30. Shammass, I., Rinott, S., Berkovitz, A., Schley, R. & Steinhauer, J. Phonon dispersion relation of an atomic Bose-Einstein condensate. *Phys. Rev. Lett.* **109**, 195301 (2012).

31. Schley, R., Berkovitz, A., Rinott, S., Shammass, I., Blumkin, A. & Steinhauer, J. Planck Distribution of Phonons in a Bose-Einstein Condensate. *Phys. Rev. Lett.* **111**, 055301 (2013).

32. Steinhauer, J. Observation of self-amplifying Hawking radiation in an analogue black-hole laser *Nature Phys.* **10**, 864 (2014).

33. Philbin, T. G., Kuklewicz, C., Robertson, S., Hill, S., König, F. & Leonhardt, U. Fiber-optical analog of the event horizon. *Science* **319**, 1367-1370 (2008).

34. Belgiorno, F., Cacciatori, S. L., Clerici, M., Gorini, V., Ortenzi, G., Rizzi, L., Rubino, E., Sala, V. G. & Faccio, D. Hawking Radiation from Ultrashort Laser Pulse Filaments. *Phys. Rev. Lett.* **105**, 203901 (2010).

35. Unruh, W. & Schützhold, R. Hawking radiation from ''phase horizons'' in laser filaments? *Phys. Rev. D* **86**, 064006 (2012).

36. Liberati, S., Prain, A. & Visser, M. Quantum vacuum radiation in optical glass. *Phys. Rev. D* **85**, 084014 (2012).

37. Nguyen, H. S., Gerace, D., Carusotto, I., Sanvitto, D. Galopin, E. Lemaître, A., Sagnes, I., Bloch, J. & Amo, A. Acoustic Black Hole in a Stationary Hydrodynamic Flow of Microcavity Polaritons. *Phys. Rev. Lett.* **114**, 036402 (2015).

38. Weinfurtner, S., Tedford, E. W., Penrice, M. C. J., Unruh, W. G. & Lawrence, G. A. Measurement of stimulated Hawking emission in an analogue system. *Phys. Rev. Lett.* **106**, 021302 (2011).

39. Rousseaux, G., Mathis, C., Maïssa, P., Philbin, T. G. & Leonhardt, U. Observation of negative-frequency waves in a water tank: a classical analogue to the Hawking effect? *New J. Phys.* **10**, 053015 (2008).

40. Parentani, R. From vacuum fluctuations across an event horizon to long distance correlations. *Phys. Rev. D* **82**, 025008 (2010).

41. Pines, D. & Nozières, Ph. *The Theory of Quantum Liquids*, Vol. I, Section 2.1 (Addison-Wesley, Reading, MA, 1988).

42. Nozières, Ph. & Pines, D. *The Theory of Quantum Liquids*, Vol. II, Section 3.1 (Addison-Wesley, Reading, MA, 1990).

43. Pitaevskii, L. & Stringari, S. *Bose-Einstein Condensation*, Section 12.9 (Oxford University Press, Oxford, 2003).